\documentclass[prb,twocolumn,superscriptaddress,aps,amsmath,amsfonts,notitlepage,showpacs]{revtex4-1}
\usepackage{amsmath,amsfonts,amssymb,dsfont,graphicx,psfrag,bm,hyperref}
\graphicspath{{/}}

\usepackage{color}

\def\be{\begin{equation}}
\def\ee{\end{equation}}
\def\bea{\begin{eqnarray}}
\def\eea{\end{eqnarray}}
\def\bpm{\begin{pmatrix}}
\def\epm{\end{pmatrix}}

\newcommand{\ve}{\varepsilon}

\begin{document}
\title{Cyclotron resonance inside the Mott gap: a fingerprint of emergent neutral fermions}
\author{Peng Rao} 
\author{Inti Sodemann}
\affiliation{Max-Planck Institute for the Physics of Complex Systems, 01187 Dresden, Germany}
\date{\today}
\begin{abstract}
A major obstacle to identify exotic quantum phases of matter featuring spin-charge separation above one-dimension is the lack of tailored probes allowing to establish their presence in correlated materials. Here we propose an optoelectronic response that could allow to pinpoint the presence of certain spin-charge separated states with emergent neutral gapless fermions in two and three-dimensional materials. We show that even though these states behave like insulators under static electric fields, they can display clear cyclotron resonance peaks in their light absorption spectrum under static magnetic fields, but typically the principal Kohn mode will be missing in comparison to ordinary metals. This distinctive phenomena could be tested in materials such as triangular lattice organics, three-dimensional mixed valence insulators YbB$_{12}$ and SmB$_6$, and transition metal dichalcogenides 1T-TaS$_2$ and 1T-TaSe$_2$.

\end{abstract}

\maketitle

\section{Introduction}
An insulator is a state of matter in which the conductivity approaches zero as the temperature and the frequency vanish. Clean band insulators have the stronger property that the real part of the conductivity is strictly zero below an optical band gap. This implies that light can propagate in band insulators without being absorbed at low frequencies. However, certain clean spin-charge separated insulating phases of matter can have the stunning property that there is a finite real conductivity that vanishes as a power of the frequency and hence there is always light absorption at low frequencies in such states~\cite{NgLee2007}.

\begin{figure}
	\includegraphics*[width=\linewidth]{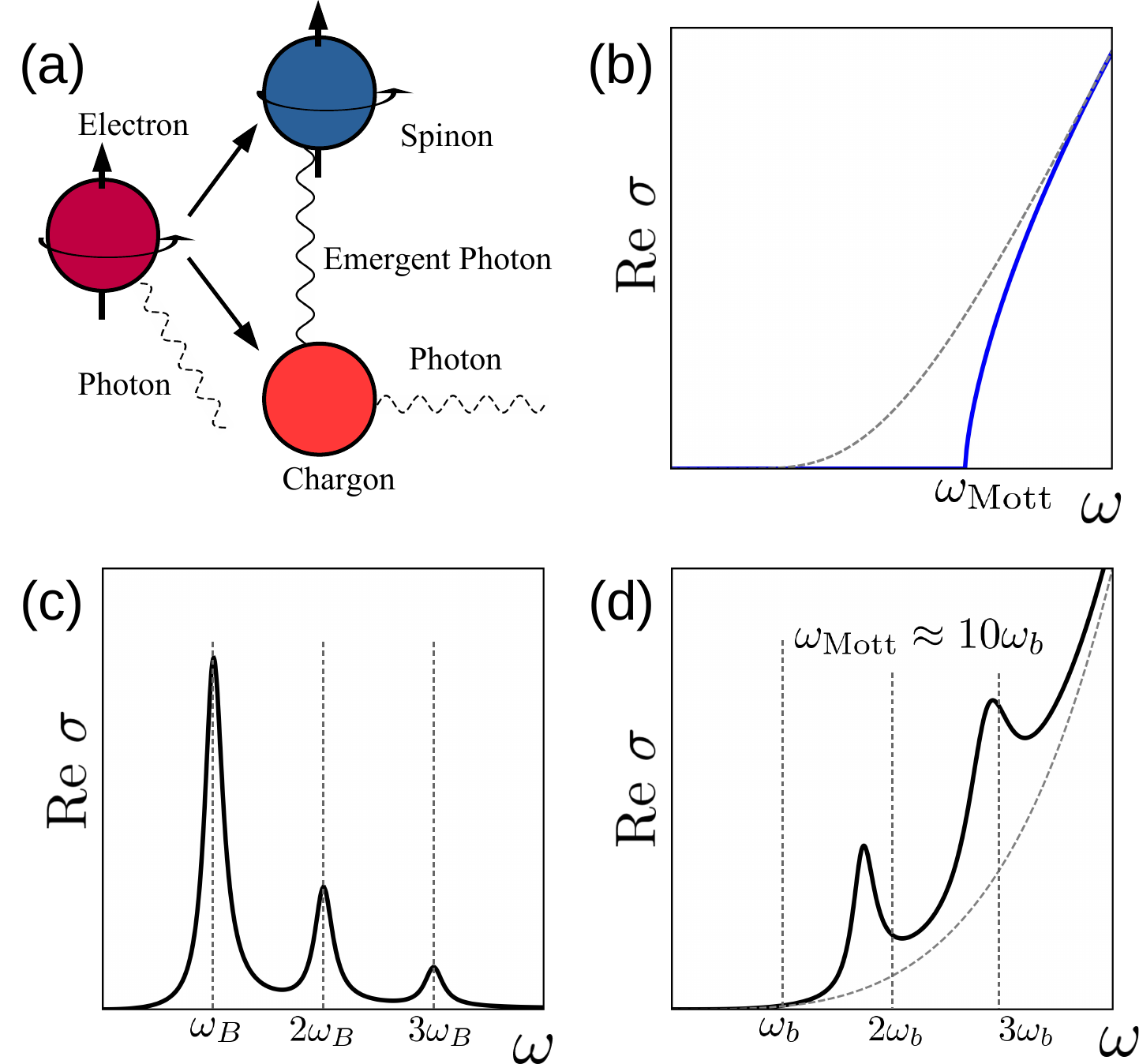}%
	\caption{\label{Fig:Schematic} (a) Depiction of the spin-charge separation pattern in the spinon fermi surface state. (b) Conductivity of clean (solid blue) and disordered (dashed) band insulators at zero temperature. (c) Conductivity of a metal in a magnetic field with parameters $\ve_F=5, \omega_{\text{Mott}}=10, \omega_B=\omega_b=1$. The weight of the $j$-th harmonic is $D_j$ with $D_1=\ve_F/\pi = 3D_2=9D_3$. (d) Conductivity of a spinon fermi surface state with (solid line) and without (dashed) magnetic field with the same band paramters as the metal in (c). There is one less low-frequency peak compared to (c).
	}
\end{figure}

This behavior occurs hand-in-hand with a pattern of spin-charge separation in which the electron ``breaks" into two pieces: a neutral fermion that carries the spin, known as the ``spinon", and a charged spinless boson, known as the ``holon" or ``chargon"~\cite{Baskaran1988}. The chargons are gapped while the spinons remain in a gapless liquid state. The spinon and the chargon carry opposite charges under an accompanying emergent U(1) gauge field, as depicted in Fig.\ref{Fig:Schematic}(a). These exotic states have been advocated to be present in triangular lattice organic materials~\cite{Zhou2017,Motrunich2005,Lee2005,Sheng2009,Lai2010,Block2011,Mishmash2015,Savary2016} $\kappa$-(ET)$_2$Cu$_2$(CN)$_3$~\cite{Shimizu2003,Yamashita2008}, EtMe$_3$Sb[Pd(dmit)$_2$]$_2$~\cite{Itou2008,Yamashita2010,Yamashita2011} and $\kappa$-H$_3$(Cat-EDT-TTF)$_2$~\cite{Isono2014}, and Herberthsmithite~\cite{Helton2007}. More recently this state has also been advocated to be present in 1T-TaS$_2$~\cite{Law2017,He2018} and a closely related state, known as the composite exciton Fermi liquid (CEFL), has been advocated to be present in correlated mixed valence materials like SmB$_6$~\cite{Chowdhury2018,Inti2018} and YbB$_{12}$~\cite{Liu2018,Sato2019,Xiang2018} allowing to understand some of their puzzling behavior~\cite{Li2014,Tan2015,Flachbart2006,Laurita2016}. This state has also been recently proposed in YbMgGaO$_4$~\cite{Shen2016,Li2016,Li2017,Li12017} although several subsequent studies have challenged this proposal~\cite{Zhu2017,Parker2018,Iaconis2018,Kimchi2018}.

Several of these materials are insulators under static electric fields but display subgap optical conductivity~\cite{Kezsmarki2006,Pinteric2014,Pilon2013, Laurita2016,Laurita2018,Pustogow2018} which could arise from spin-charge separation~\cite{NgLee2007,Potter2013,Chowdhury2018}. However, one of the common competing hypothesis  with the spin-charge separation scenario is the existence of inhomogeneous states that could be spontaneously formed or induced by disorder~\cite{Kawamoto2004,Nakajima2012,Pinteric2014,Padmalekha2015}. Such disorder can smear the sharp optical gap that would otherwise be present in a clean band insulator making it hard to distinguish this situation from the ideal subgap optical absorption that accompanies the spin-charge separated state, as depicted in Fig.\ref{Fig:Schematic}(b). Such disorder could also account for a linear temperature dependence in specific heat without an accompanying linear temperature dependence in heat conductivity, as recent measurements~\cite{Ni2019,Bourgeois2019} have found in organic materials.

In this article  we study an opto-electronic response of these spin-charge separated states that could sharply distinguish them from conventional insulators and metals, providing a fingerprint for their presence. We will argue that the spectrum of optical absorption of these states in the presence of magnetic fields can display clear peaks below the Mott scale that resemble the cyclotron resonance peaks of metals, allowing to distinguish them from conventional insulators, but also that among these peaks the principal Kohn-mode harmonic is absent allowing also to distinguish them from conventional metals, as depicted in Fig.\ref{Fig:Schematic}(d). This scenario could be systematically tested in materials that are not far from the metal-insulator transition~\cite{Florens2004,Senthil2008}, especially so if the transition can be tuned by experimentally amenable knobs such as pressure~\cite{Kurosaki2005,Itou2017} or magnetic fields~\cite{Xiang2018}.


\section{General Setting}

We are interested in phenomena arising in the spinon Fermi surface at energies that are much smaller than the insulating or Mott scale, which can be viewed as the chargon gap~\cite{Florens2004}. At such scales the effective field theory reduces to that of a Fermi surface of spinons minimally coupled to an emergent U(1) gauge field, denoted by $a$~\cite{Ioffe1989,Lee1992,Kim1994,Senthil2008}:

\begin{equation}\label{Eqn:TotalLagrangian}
\mathcal{L} = \mathcal{ L}_\Psi (p_\mu-a_\mu) + \frac{\kappa}{2} (E-e)^2-\frac{1}{2\mu} (B-b)^2,
\end{equation}
here $\mathcal{L}_\Psi$ is the kinetic term describing the band dispersion of the spinons, $e$ and $b$ are the electric and magnetic fields associated with the emergent gauge field $a$, and $E$ and $B$ are the physical probe electric and magnetic fields. This Lagrangian suggests that the emergent $b$ and $e$ fields tend to track the physical $E$ and $B$ fields.

In fact, an external magnetic field $B$ can induce an average internal magnetic field $b$~\cite{Motrunich2006,Katsura2010,Lai12010,Gao2019,Inti2018,Chowdhury2018}:
\[
b \approx \frac{1}{1+\mu \chi_\Psi} B.
\]
Here $\chi_\Psi$ is the Landau diamagnetic coefficient of the spinons given by $1/12\pi m_\Psi$ for a parabolic dispersion with mass $m_\Psi$~\cite{Inti2018}. Therefore, in the presence of an external magnetic field, the spinons experience cyclotron quantization leading to magnetization oscillations analogous to the de-Haas van Alphen effect~\cite{Motrunich2006,Inti2018,Chowdhury2018} which could be the mechanism behind the quantum oscillations seen in SmB$_6$ and YbB$_{12}$~\cite{Xiang2018,Li2014,Tan2015}, a phenomenon of much recent theoretical debate~\cite{Knolle2015,Knolle2017,Zhang2016,Baskaran2015,Erten2017,Grubinskas2018,Pal2017,Pal12017,Ram2017}.

The central question that we address in the present study is how such cyclotron quantization impacts the optical absorption spectrum of the spinon fermi surface state and, specifically, whether an analogue of the cyclotron resonance phenomenon observed in metals is also present. The optical absorption is determined by the real part of the conductivity $\sigma(\omega)$. 

As demonstrated in the seminal work of Ioffe and Larkin~\cite{Ioffe1989}, the net physical conductivity of this state can then be written as:
\begin{equation}\label{Eqn:ILRule}
\sigma^{-1} = \sigma^{-1}_\varphi + \sigma^{-1}_\Psi.
\end{equation}
$\sigma_\varphi$ and $\sigma_\Psi$ are the chargon and spinon conductivity tensors. Because the chargons are gapped, their conductivity for frequencies below the Mott scale is that of ordinary dielectric insulators: $\sigma^{ij}_\varphi = -i\omega  \kappa \delta^{ij}$, where $i, j$ are spatial indices. The problem therefore reduces to computing the spinon conductivity $\sigma_\Psi$ in the presence of $b$. The conductivities can be expressed as a sum of multiple resonances:
\begin{equation}\label{Eqn:SpinonConductivity}
\sigma_{xx}(\omega) = i\sum_{j=1} \frac{\omega D_j}{\omega^2-\omega_j^2},~\sigma_{xy}(\omega) = -\sum_{j=1} \frac{\omega_j D_j^{'}}{\omega^2-\omega_j^2}.
\end{equation}
Here $D_j$ and $D'_j$ are quantities with units of Drude weight (density divided by mass) measuring the strength of the $j$-th resonance at frequency $\omega_j$. The frequency $\omega$ has a finite imaginary part $\omega+i\Gamma$ that accounts for the width of the $j$-th resonance and that is implicit in Eqn.(\ref{Eqn:SpinonConductivity}).

The cyclotron spectrum depends on the details of the fermi surface geometry and energetics, and in general can be fairly complex. The simplest case is that of a parabolic band, in which the cyclotron resonances consists of a single peak at the cyclotron frequency, namely, $D_1=D_1'=n/m_\Psi$, and $\omega_1=b/m_\psi$, while $D_{j\geq2}=0$, known as the Kohn mode~\cite{Kohn1961}. In a parabolic band metal the Kohn mode evolves continously from the Drude peak at zero field and inherits the complete Drude weight. Remarkably, in this case one can show that the spinon fermi surface state has no cyclotron resonance peaks with frequencies comparable to the spinon cyclotron frequency $\omega_1=b/m_\Psi$. The absence of a low energy Kohn mode is a direct consequence of the strong coupling of spinons to a gauge field which transforms the Kohn mode into a plasma-like excitation with high energies of the order $\omega_{\text{Mott}}=(D/\kappa)^{1/2}$, as we explain in more detail in Appendix \ref{Appendix:Model} using an intuitive equation of motion description. This plasma scale can be interpreted as a proxy for the Mott scale within our effective low energy description. These modes will be essentially impossible to discern experimentally in $q=0$ probes such as the optical conductivity because they essentially lead to a small splitting of the wide Mott optical lobe of the order of cyclotron frequency.

However, the key finding of our study is that when the spinon dispersion is not a simple single parabolic band, such that the spinon conductivity has $N$ distinct cyclotron resonance peaks, the physical conductivity of the spinon fermi surface state will in general display $N-1$ resonances at low energy frequencies of the order of the cyclotron energy of the spinons. Moreover, in the special case in which the system is a spinon compensated semimetal, such as the CEFL proposed in Ref.\onlinecite{Chowdhury2018}, namely when the Fermi volume of particle-like spinon excitations equals exactly the Fermi volume of hole-like spinons, we will show that the physical conductivity displays $N-2$ low energy resonances. The observation of such characteristic resonance patterns would be a strong evidence of the presence of these kind of spin-charge separated states in correlated materials. We will first illustrate this within a simple model with two parabolic valleys that allows to transparently understand the essential properties of the cyclotron resonance in spinons and subsequently we will discuss the case of a fermi surface with C$_4$ symmetry that can be viewed as a simplified model of fermi surfaces in materials with cubic symmetry such as YbB$_{12}$ and SmB$_6$.

\section{Two-Valley Model}

The simplest model that displays a single low energy spinon cyclotron resonance peak is one in which the spinons have a band structure with two parabolic valleys around different crystal momenta that are not connected by any symmetry so that the effective masses and fermi surface volumes are generically distinct. In such case, we would have two spinon Drude weights $D_j=D_j'=n_j/m_j$, cyclotron energies $\omega_j=b/m_{j}$, and cyclotron life-times $\Gamma_j$ with $j=1,2$ labelling the valleys. Within this model one can show that (see Appendix \ref{Appendix:Model}) the spinons have a single resonance with a frequency and a width given by:
\begin{equation}\label{Eqn:EEPole}
\omega_\Psi + i\Gamma_\Psi =\frac{D^{-1}_1(\omega_1+i\Gamma_1) +D^{-1}_2 (\omega_2 +  i \Gamma_2)}{D^{-1}_1+D^{-1}_2}.
\end{equation}
The above equation implies that the physical resonance will tend to have a frequency closer to that of the cyclotron frequency of the valley with the smallest spinon Drude weight. The conductivity near the resonance can be approximated as:
\begin{multline}\label{Eqn:EETotalConductivity}
\sigma_{xx} \approx \frac{\omega^2 \bigg(\omega^2-\frac{D_2\omega_1^2+D_1\omega_2^2}{D_1+D_2}\bigg)}{\omega^{~4}_{\text{Mott}}} \times \frac{i\omega (D_1+D_2)}{(\omega+i\Gamma_\Psi)^2-\omega_\Psi^2}, \\ 
\sigma_{xy} \approx \frac{\omega^2 \bigg(\omega^2-\frac{\omega_1 D_1\omega_2^2+\omega_2D_2\omega_1^2}{D_1\omega_1+D_2\omega_2}\bigg)}{\omega^{~4}_{\text{Mott}}} \times \frac{i(\omega_1 D_1+\omega_2D_2)}{(\omega+i\Gamma_\Psi)^2-\omega_\Psi^2},
\end{multline}
where $\omega^{~2}_{\text{Mott}} = (D_1+D_2)/\kappa$. 

\begin{figure}
		\includegraphics*[width=\linewidth]{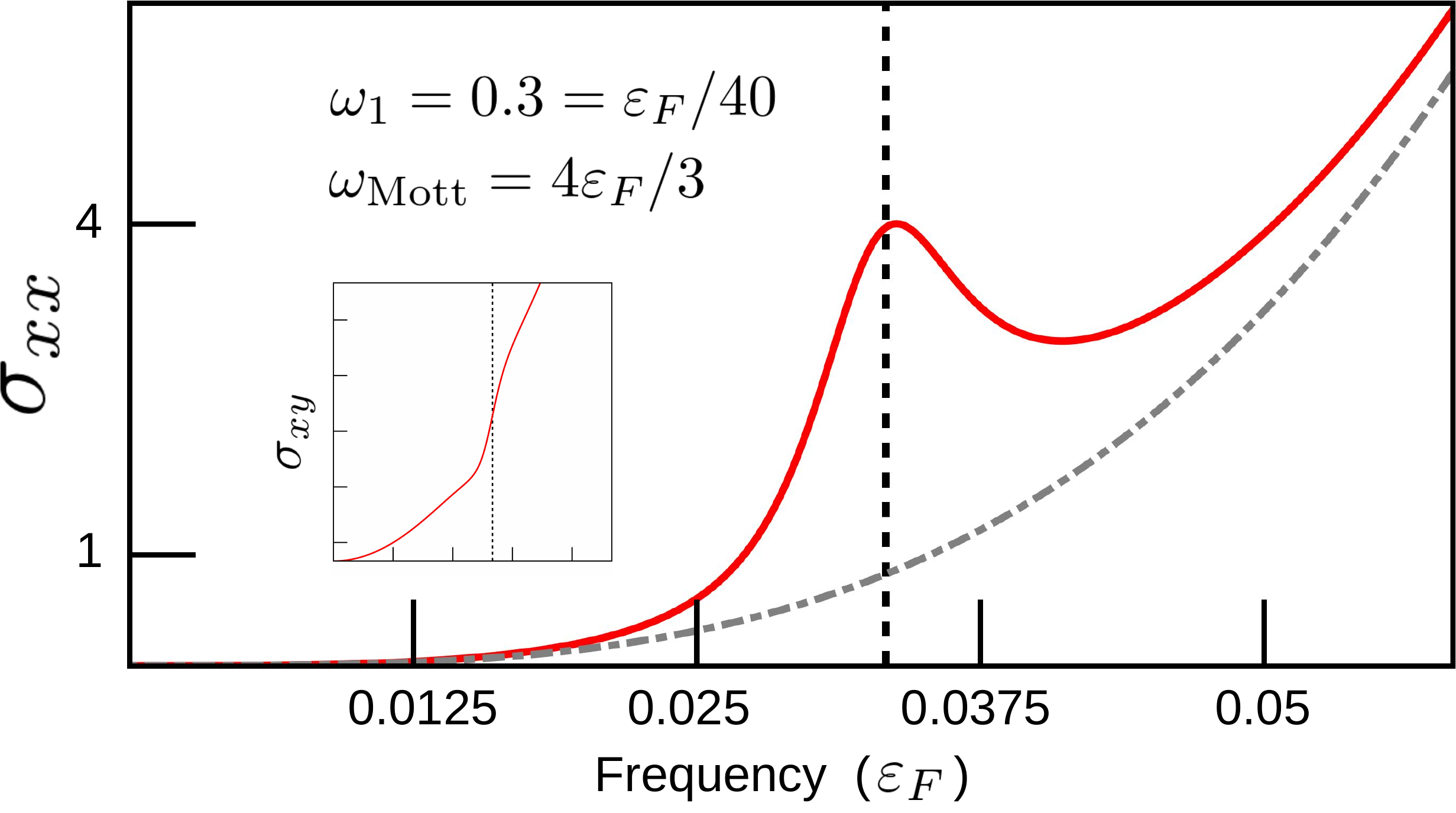}
	\caption{\label{Fig:TwoPockets} Conductivity for a two-valley model of spinons with (solid red) and without (dashed gray) magnetic field. The valleys have $\ve_F=12, D_2=2D_1=2\ve_F/\pi, \omega_1=  \ve_F/40, \omega_2=\ve_F/20,$ and $ \omega_\text{Mott}\approx 4\ve_F/3$. The vertical line gives the resonance frequency predicted by (\ref{Eqn:EEPole}) $\omega_\Psi=4\omega_1/3$. The spinon width is taken from Eqn.(\ref{Eqn:Width2} for $d=2$. The inset depicts the frequency dependence of the Hall conductivity. The conductivity is measured in units of its zero-field value at the resonance.
	}
\end{figure}

Fig.\ref{Fig:TwoPockets} illustrates behavior of the conductivity for this model. From the above equations we see that the weight of the spinon resonance decays as $\omega_{\text{Mott}}^{-4}$, and therefore the peaks will become less pronounced the deeper into the insulating Mott regime the spin liquid is. Also the above relations make transparent that the spectral weight of the resonance vanishes when the spinon cyclotron frequencies of each valley become the same: $\omega_1 \rightarrow \omega_2 $, in agreement with the discussion of the previous section according to which when the spinon conductivity has a single resonance the physical conductivity has no discernible resonances at low energies. A complementary explanation for this finding is presented in Appendix \ref{Appendix:Model} based on the equation of motion approach.

The case of two parabolic valleys, one with particle-like and the other with hole-like spinons, can be obtained from the previous formulae by changing $\omega_2 \rightarrow-\omega_2$, if we take $`2'$ to be the hole-like valley. The expression for the resonance frequency in this case is:
\begin{equation}\label{Eqn:EHPole}
\omega_\Psi = \frac{|D_2^{-1}\omega_2-D_1^{-1}\omega_1|}{D_1^{-1}+D_2^{-1}},~\Gamma_\Psi = \frac{|D_1^{-1}\Gamma_1+D_2^{-1}\Gamma_2|}{D_1^{-1}+D_2^{-1}}.
\end{equation}
The compensated limit is approached when $\omega_1/D_1\rightarrow \omega_2/D_2$, and it is clear that in this case the resonance frequency vanishes $\omega_\psi \rightarrow 0$, and also the weight of the resonance vanishes as $\omega_\Psi/\omega_{\text{Mott}}^2$. Therefore we see that generically we have a single resonance in the two valley system, except when the valleys are of particle-hole type and have exactly the same Fermi volume, in agreement with the more general rule that in the compensated case the spinon Fermi surface state will have two less resonances than those expected for ordinary metals. Detailed analytic understanding of the case with more than one resonance is cumbersome, but plotting the conductivity is straightforward by using the Ioffe-Larkin rule. As an example, Fig.\ref{Fig:Schematic}(d) illustrates the case of two physical cyclotron resonances that arises from a model in which the spinon conductivity has three resonances.

\section{$C_4$ Symmetric Model}

Even if a system has a single fermi surface it will generically display several cyclotron resonances because the dispersion will deviate from being perfectly parabolic. This is particularly true when the system has a large Fermi surface comparable with the size of the Brillouin zone, such as a half-filled single orbital Hubbard model in the triangular lattice~\cite{Zhou2017}, and thus these systems are expected to display a multiplicity of cyclotron harmonics inside their Mott gap. To illustrate this case we consider a C$_4$-symmetric dispersion for the spinons given by:
\begin{equation}\label{Eqn:Spectrum}
\varepsilon(p)=\frac{p^2}{2m}+\frac{g p^4 }{2} \cos 4\phi.
\end{equation}
Here we take magnetic field along the z-axis and $\phi$ is azimuthal angle of momentum. $g$ controls the strength of the deviation of the Fermi surface from a perfect circle which we take to be small. This model therefore serves as a semi-realistic description for systems with small spinon Fermi surfaces and cubic symmetry which might be relevant to describe the observations in YbB$_{12}$~\cite{Sato2019}. 

\begin{figure}
	\includegraphics*[width=\linewidth]{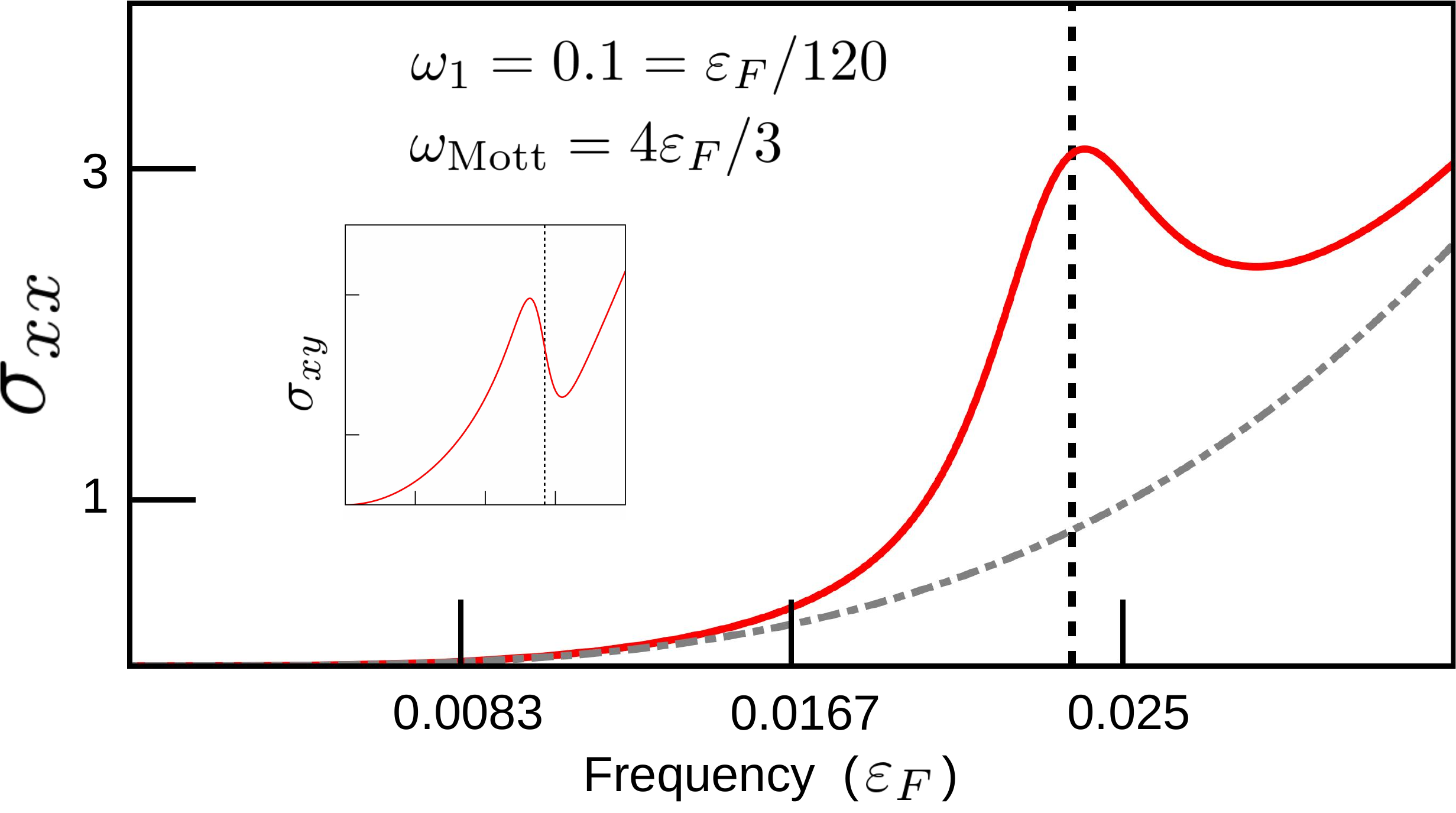}%
	\caption{\label{Fig:Anisotropy} Conductivities of the C$_4$ symmetric Fermi surface of spinons with (solid red) and without (dashed gray) magnetic field. The spinon band parameters are chosen as $\ve_F=12, D_1=\ve_F/\pi, \omega_1= \ve_F/120, \omega_2=3\omega_1, g=0.2, D_2=-D_2^{'}=g^2 D_{1}, \omega_\text{Mott}\approx 4\epsilon_F/3$. The spinon width is taken from Eqn.(\ref{Eqn:Width2} for $d=2$. The vertical line gives the resonance frequency predicted by (\ref{Eqn:EEPole}) $\omega_\Psi\approx 2.85 \omega_1$. The inset depicts the frequency dependence of the Hall conductivity. The conductivity is
	measured in units of its zero-field value at the resonance.
	}
\end{figure}

To order $g^2$ one encounters a spinon conductivity with two resonances at $\omega_b$ and $3\omega_b$. Details of calculations are given in Appendix \ref{Appendix:Anisotropy}. Drude weights of the principal harmonic receive small perturbative corrections and satisfy $D_1=D_1^{'}$. The third harmonic is given in terms of spinon number density $n$: $D_2 = -D^{'}_2=36\pi^2g^2 n^3b$ in the quasi-classical limit. This is formally the same as a system with particle and hole valleys with the Drude weights $D_1, D_2$. Therefore the conductivity is given by (\ref{Eqn:EHPole}) with $\omega_1=\omega_b$ and $\omega_2=3\omega_b$. Given that $D_2\ll D_1$, from (\ref{Eqn:EHPole}) the physical conductivity is expected to display a cyclotron resonance peak near $3\omega_b$ while the peak at $\omega_b$ is absent. The residue of the conductivity peak is related to the spinon Drude weights as follows:
\[
\text{res}~\sigma_{xx}=36\bigg( \frac{\omega_b}{\omega_{\text{Mott}}}\bigg)^4 D_{2},~\text{res}~\sigma_{xy}=72\bigg( \frac{\omega_b}{\omega_{\text{Mott}}}\bigg)^4 D_{2}.
\]
Due to the $(\omega_b/\omega_{\text{Mott}})^4$ suppression we see that the cyclotron peaks in the spinon fermi surface state typically has a smaller weight than an ordinary metal, but the resonance should be visible provided the system is not far from the Mott transition, as illustrated in Fig.\ref{Fig:Anisotropy}.

\section{Effects of Gauge Field Fluctuations}

To ensure that the spinon cyclotron resonance peaks are discernible, it is crucial to verify that their width is smaller than the typical frequency at which the resonances are expected. This is particularly relevant because the spinons are not sharp quasiparticles, since the quasiparticle lifetime scales as $\varepsilon^{d/3}$ for $d=2,3$ dimensions~\cite{Lee1992,Kim1994,Polchinski1994,Podolsky2009,Lee2018}. However, the lifetime that determines the conductivity is not the quasi-particle but the transport lifetime, which determines the rate at which the spinons lose momentum through the internal radiation of emergent photons. This process gives rise to a form of frictional force on the spinon motion through the radiation reaction that accompanies emergent photon emission except that the propagation of the emergent photon occurs self-consistently in the spinon fermi surface medium and as a result it is subject to a strong Landau damping~\cite{Lee1992}. The transport life-time can be estimated to be~\cite{Lee1992,Zhou2013}:
\begin{equation}\label{Eqn:Width}
\Gamma_d = C_d \omega \bigg(\frac{\omega}{\Lambda}\bigg)^{(d-1)/3},
\end{equation}
where $\Lambda$ is an energy scale that is of the order of the spinon Fermi energy in the Mott insulating regime, and $C_d$ is a dimensionless constant of order unity. Since at low frequencies $\Gamma \ll \omega$, we therefore expect the spinon cyclotron resonance peaks to be discernible in two and three dimensions. Specific estimates for this transport lifetime are presented in Appendix \ref{Appendix:Fluctuations}. Throughout the plots we have used the two dimensional estimate in Eqn.(\ref{Eqn:Width2}, where the gauge fluctuations are more severe. The resonances are therefore expected to be sharper in three-dimensions. In the presence of static disorder the width might be limited by the transport time $\tau_0$ associated with elastic collisions, and one expects the effective width to be given by a Matthiessen type rule $\Gamma_{\text{eff}}=1/\tau_0+\Gamma(\omega)$.

\section{Conclusions}

We have shown that in spite of behaving like an insulator in response to static electric fields, the spinon fermi surface state can display clear resonant peaks in its frequency dependent conductivity in the presence of magnetic fields resembling the cyclotron resonance of metals. The observation of this effect requires certain conditions. In particular, the system should not be too strongly insulating since the residues of peaks decreases as $1/\omega_{\text{Mott}}^{4}$. However if this phenomena is observed it would be very hard to reconcile with a conventional insulator with disorder as the states deep inside the tails of the optical gap tend to be the most localized ones and hence the least susceptible to developing cyclotron motion in response to magnetic fields. In this sense this effect serves as a fingerprint of the gapless emergent neutral spinons coupled to an emergent U(1) gauge field. In Appendix \ref{Appendix:Estimates}, we list order-of-magnitude estimates of the energy scales relevant for cyclotron resonance in several candidate materials. Finally, we remark that although we have focused on the cyclotron resonance for spinons with a Fermi surface, similar phenomena is expected to occur for spinons with nodal dispersions such as in Dirac spin liquids, provided they are also coupled to a U(1) gauge field.

\medskip

We thank D. Chowdhury, T. Senthil and P. A. Lee for previous collaborations and discussions that inspired this work, and J. Falson and Y. Matsuda for valuable discussions and correspondence.

\appendix

\section{Equation of motion for two-valley spinons}\label{Appendix:Model}

The specific formulas and plots in the main text are based on the Ioffe-Larkin rule which is a powerful way to estimate the electromagnetic response of parton constructions. In this section we present a simplified model that allows to gain intuition on the dynamics of spinons and on why the Kohn mode is absent in their cyclotron resonance. The agreement between the Ioffe-Larkin and rule and our simple model is primarily due to that we are computing $q=0$ correlation functions which are controlled by the globally spatially averaged behavior the liquids.

The physical picture behind the cyclotron resonance in conductivity below the optical gap can be understood by considering the classical motion of spinons in a dielectric. Their equations of motion must be solved simultaneously with those of the gauge field which is also a dynamic degree of freedom. We retain here only the global spatial average of the electric field as the only dynamical variable of the gauge fields, and hence write an the following effective Lagrangian:
\begin{equation}\label{Eqn:EffectiveLagrangian}
\mathcal{L} =   \mathcal{L}_{\Psi_1}  +  \mathcal{L}_{\Psi_2} +\frac{\kappa}{2} (\dot{{\bm a}}+{\bm E})^2.
\end{equation}
The first two terms are Lagrangians for the two valleys of spinons including their couplings to the gauge field. The second term is the kinetic energy of the emergent electric field consistent with Eqn.(\ref{Eqn:TotalLagrangian}) in the main text. The Euler-Lagrange equations are:
\begin{eqnarray}\label{Eqn:ELEquations}
\kappa(\dot{{\bm a}}+{\bm E}) &=&  \sum_i^{N_1} {\bm r}_{1i} + \sum_i^{N_2} {\bm r}_{2i},\\
m_1\dot{{\bm v}}_{1i}  + {\bm v}_{1i} \times{\bm b}&=& -\dot{{\bm a}},~i = 1,...,~N_1,\\
m_2\dot{{\bm v}}_{2i}  +{\bm v}_{2i}\times{\bm b}&=& - \dot{{\bm a}},~i=1,...,~N_2.
\end{eqnarray}
${\bm r}_{1,2 i}$ is the position vector of the i-th spinon in valley $1,2$. Summing over $i$ from $1$ to $N_1$ in the second and $1$ to $N_2$ in the third equations and substituting for $\dot{{\bm a}}$ from the first, we obtain two equations for the spinon centers of mass:
\begin{eqnarray}\label{Eqn:ELEquations1}
m_1\dot{{\bm V}_1}  + {\bm V}_1\times{\bm b} +(N_1 {\bm R}_1 + N_2 {\bm R}_2)/\kappa&=&{\bm E}, \\
m_2\dot{{\bm V}_2} + {\bm V}_2\times{\bm b} +(N_2 {\bm R}_2 + N_2 {\bm R}_1)/\kappa &=&{\bm E},
\end{eqnarray}
where ${\bm R}_1 = \sum_i^{N_1} {\bm r}_{1i}/N_1$ and ${\bm R}_2=\sum_i^{N_2} {\bm r}_{2i}/N_2$. The normal modes associated with Eqn.(\ref{Eqn:ELEquations1}) can be computed exactly because these are linear ordinary differential equations. One encounters three resonances. Two of them are high energy resonances close to the Mott scale and the other is the low frequency resonance. They are given by:
\begin{eqnarray}
\omega_{\text{Mott}1,2}=[(D_1+D_2)/\kappa]^{1/2}\pm \frac{(D_1^2+D_2^2)}{2D_2(D_1+D_2)}\omega_{1},\\
\omega_\Psi =\frac{D^{-1}_1\omega_1 +D^{-1}_2 \omega_2 }{D^{-1}_1+D^{-1}_2},
\end{eqnarray}
where $\omega_{1,2}=b/m_{1,2}$.  As we see above, the low energy resonant frequency coincide exactly with the main text result Eqn.(\ref{Eqn:EEPole}) which is derived using the Ioffe-Larkin rule Eqn.(\ref{Eqn:ILRule}) in the main text. We also see in Eqs.(\ref{Eqn:ELEquations1}) that the coupling of the electric field tends to produce a pinning term for the center of mass motion of the spinons, as if it was coupled to a ``spring" of constant $1/\kappa$, and the Mott scale appears then as the frequency of resonance of such center of mass motion. This pinning of the center of mass motion is responsible for the absence of the Kohn mode resonance relative to the case of ordinary metals.

Let us now investigate in more detail the case of $m_1=m_2$ and $ \omega_1=\omega_2$, in which there is no low energy cyclotron resonance. Now Eqn.(\ref{Eqn:ELEquations1}) still predict low energy cyclotron resonance modes at $\omega=\pm\omega_1$, so it is not immediately clear why such resonances disappear from the cyclotron spectrum. The key to understand this is to analyze the how such modes couple to the physical electric current. As we will shortly see such modes are still present but they become electrically neutral in the sense that they are not excited by the oscillatory physical electric field. To see this we first identify the physical electric current associated with the Lagrangian in Eq.(\ref{Eqn:EffectiveLagrangian}):
\begin{equation}\label{Eqn:Current}
{\bm j} = \delta \mathcal{L}/\delta {\bm A} =N_1 {\bm V}_1 + N_2 {\bm V}_2,
\end{equation}
where we have used Eqn.(\ref{Eqn:ELEquations}) in deriving the last equality. The velocity eigenmodes to (\ref{Eqn:ELEquations1}) can be written in the 4-dimensional $({\bm R}_1,{\bm R}_2)$ vector space as ${\bm V}=({\bm V}_1,{\bm V}_2)$ where
\[
{\bm V}(t) = {\bm V}_0 \exp (-i\omega t).
\]
In particular, for the two low energy resonances $\pm \omega_1$ at $m_1=m_2$, ${\bm V}_0$ is proportional to
\[
(-iN_2,-N_2,iN_1,N_1),~(iN_2,-N_2,-iN_1,N_1).
\]
These are exact clock-wise and anti-clockwise circular motions of the two fermion centers of mass. The current (\ref{Eqn:Current}) mode can be written in the same 4-dimensional space as:
\[
{\bm j} = (N_1,N_1,N_2,N_2).
\]
The overlap between ${\bm j}$ and the two low energy resonances ${\bm j}.{\bm V}_0$ is zero. Thus cyclotron resonance modes are not excited by ${\bm E}$. Repeating the above derivations for one type of spinons only gives only two high frequency Mott peaks split by $\omega_b$. 

For electron and hole excitations, we obtain similar equations to (\ref{Eqn:ELEquations1}):
\begin{eqnarray}\label{Eqn:ELEquations2}
m_e\dot{{\bm V}}_e  + {\bm V}_e\times{\bm b} +(N_e {\bm R}_e - N_h {\bm R}_h)/\kappa&=&{\bm E}, \\
m_h\dot{{\bm V}}_h - {\bm V}_h\times{\bm b} +(N_h {\bm R}_h - N_e {\bm R}_e)/\kappa &=& -{\bm E}.
\end{eqnarray} 
The low energy eigenvalue $\omega_\Psi$ coincides with (\ref{Eqn:EHPole}) in the main text in the small $\kappa$ limit. At $N_e=N_h=N$, $\omega_\Psi=0$ and the 4D eigenmodes ${\bm V}=({\bm V}_{0e},{\bm V}_{0h}) \exp (-i\omega t)$ are proportional to:
\begin{eqnarray}
&&\omega_{\text{Mott}}: (-1,0,1,0),~(0,-1,0,1); \\
&&\omega_\Psi: (1,0,1,0),~(0,1,0,1),
\end{eqnarray}
while the physical current ${\bm j} = N({\bm V}_e-{\bm V}_h)$ is:
\[
{\bm j}=N(1,1,-1,-1).
\]
The Mott peak modes couple coherently to the physical current while the low energy mode, which corresponds to translation of the system as a whole, has no overlap with the current mode: the cyclotron resonance modes are charge neutral and not excited by ${\bm E}$.

\section{Derivation of the conductivity tensor}\label{Appendix:Conductivity}
First, we show that if the spinon resistivity tensor $\rho$ has one pole in frequency at  $\omega=\omega_0\ll \omega_{\text{Mott}}$, $\sigma$ has the same pole. Let $\rho_{xx}$ and $\rho_{xy}$ have the form:
\begin{equation}\label{Eqn:SpinonResistivity}
\rho_{xx} =i \frac{f(\omega)}{\omega-\omega_0},~\rho_{xy}= \frac{g(\omega)}{\omega-\omega_0}.
\end{equation}
In the range $\omega\sim \omega_0\ll \omega_{\text{Mott}}$, the small $\kappa$ limit can be taken in Eqn.(\ref{Eqn:ILRule}) in the main text. The relevant pole term in $\sigma$ turns out to be:
\begin{equation}\label{Eqn:ILConductivity}
\sigma_{xx} = -i\kappa^2\frac{\omega^2 f(\omega)}{\omega-\omega_0},~\sigma_{xy}=\kappa^2\frac{\omega^2 g(\omega)}{\omega-\omega_0}.
\end{equation}
In the main text, Eqns.(\ref{Eqn:EEPole}) and (\ref{Eqn:EETotalConductivity}) are then obtained by directly inverting the spinon conductivity (\ref{Eqn:SpinonConductivity}) with two valleys in total.

\section{Conductivity for systems with weak square-symmetric anisotropy}\label{Appendix:Anisotropy}
In a uniform emergent magnetic field ${\bm b}$, in-plane quasi-momentum operators in the free electron Hamiltonian $H_0={\bm p}^2/2m$ are substituted by creation and annihilation operators on Landau-Level (LL) indices:
\begin{eqnarray}
p_x \rightarrow \frac{1}{\sqrt{2}l_b}(a+a^\dagger), ~ p_y \rightarrow \frac{i}{\sqrt{2}l_b}(a^\dagger-a),\\
H_0 = \hbar \omega_b \bigg(a^\dagger a + \frac{1}{2}\bigg),~E_n = \hbar \omega_b \bigg(n+ \frac{1}{2}\bigg)\\
a |n\rangle = \sqrt{n} |n-1 \rangle.
\end{eqnarray}
$m$ is the eletron mass, $\omega_b=b/m$ cyclotron frequency and $l_b = \surd(\hbar/b)$ the magnetic length. $|n \rangle$ is the $n$-th energy eigenstate of $H_0$ where the intra-LL index has been omitted.

The Hamiltonian given by Eqn.(\ref{Eqn:Spectrum}) in the main text then becomes
\begin{equation}	
H=H_0+V,~V = \frac{g}{l_b^4} (a^{\dagger 4}+a^4).
\end{equation}
$g$ is the parameter controlling the Fermi surface anisotropy that we will take to be small. Correspondingly, the current operators also receive perturbation correction to first order:
\begin{eqnarray}
j_x &=& \frac{1}{\sqrt{2}ml_b}(a+a^\dagger) + \frac{2\sqrt{2}g}{l_b^3} (a^3+a^{\dagger3}),\\
j_y &=&\frac{1}{\sqrt{2}ml_b}(a^\dagger-a)  +\frac{2\sqrt{2}g}{l_b^3} i g (a^3-a^{\dagger3}).
\end{eqnarray}
Due to C$_4$ symmetry, $\sigma$ only has two independent components $\sigma_{xx}, \sigma_{xy}$ while $\sigma_{yy}=\sigma_{xx}$ and $\sigma_{yx}=-\sigma_{xy}$. We compute the linear response conductivity tensor at zero temperature using Kubo's formula:
\begin{multline}\label{Eqn:Kubo}
\sigma_{xx}(\omega) = -\frac{i}{\omega} \sum_{m>n}[n(\varepsilon_m)-n(\varepsilon_n)] |\langle n | j_x | m \rangle|^2 \frac{2\omega_{mn}}{\omega^2-\omega^2_{nm}},\\
\sigma_{xy}(\omega)=\sum_{m>n}[n(\varepsilon_m)-n(\varepsilon_n)] \langle n | j_x | m \rangle \langle m | j_y | n \rangle \frac{2}{\omega^2-\omega^2_{nm}}.
\end{multline}
$\varepsilon_n$ is the n-th energy level, $\omega_{mn}=\varepsilon_m-\varepsilon_n$ and $n(\varepsilon)$ is the Fermi-distribution. At $T=0$, summation is taken over all transitions from filled to unfilled LLs. We denote by $N$ the last filled LL index. The full conductivity tensor satisfies
\begin{equation} \label{Eqn:Constraints}
\sigma_{xx}(0)=0,~\sigma_{xy}(0) = n /2\pi
\end{equation}
due to that completely filled LLs are gapped and the quantisation of Hall conductivity. Here $n=(N+1)N_\phi$ is the particle number density and $N_\phi=b/2\pi$ is the LL degeneracy. Perturbation theory must be done carefully by verifying that these constraints are not violated at any order. This allows us to transform $\sigma_{xx}$ into a more convenient form as follows. $\sigma_{xx}$ can be written in the form
\begin{equation}\label{Eqn:Kubo1}
\sigma_{xx}(\omega) = \frac{i}{\omega} \bigg(\sum_{j=1} \frac{\omega_j^2D_j}{\omega^2-\omega_j^2} +D_0\bigg).
\end{equation}
$D_0=n/m$ is due to the linear vector potential part in the current operator and equal to the Drude weight for parabolic spinons. $\sigma_{xx}(0)=0$ gives $\sum_j D_j = D_0$. Substituting for $D_0$ in (\ref{Eqn:Kubo1}) gives
\begin{equation}\label{Eqn:Kubo2}
\sigma_{xx}(\omega) = i\sum_{j=1} \frac{\omega D_j}{\omega^2-\omega_j^2}.
\end{equation}

To satisfy the second equality of (\ref{Eqn:Constraints}), in (\ref{Eqn:Kubo}) one must include perturbation corrections to free-electron LLs and energy levels. The conductivity would then have $n$ poles centered around $\pm n\omega_b$ instead of one since the energy spacings are not uniform. This difference can be neglected in the semiclassical limit in which the Landau level spacing becomes uniform near the Fermi energy, and the higher cyclotron harmonics can therefore be approximated as single poles. Taking above into account, perturbation calculations show that the lowest order spinon conductivity corrections are proportional to $g^2$, with cyclotron poles at $\pm \omega_b,~\pm3\omega_b$ respectively.  In particular, the $g^2$ order perturbative correction to 'Drude weights' of the principle harmonic is:
\[
\delta D_1 =\delta  D_1^{'} = 21n g^2(N^2+2N+2)/l_b^6.
\]
This is simply a small change compared to the parabolic band Drude weight and can be safely neglected. Near the $\pm3\omega_b$ poles, perturbation corrections to $\pm \omega_b$ poles can be neglected and we write the spinon conductivity tensor:
\begin{eqnarray}\label{Eqn:Kubo3}
&&\sigma_{xx} = i\frac{\omega D_2}{\omega^2-9\omega^2_b} + \sigma^0_{xx}, ~\sigma^0_{xx} = \frac{i\omega D_0}{\omega^2-\omega^2_b},\\
&& \sigma_{xy}  = -\frac{3\omega_b D^{'}_2}{\omega^2-9\omega^2_b}+ \sigma^0_{xy},~\sigma^{0}_{xy}=-\frac{\omega_bD_0}{\omega^2-\omega^2_b},
\end{eqnarray}
where
\[
D_2=-D^{'}_2=9g^2n (N^2+2N+2)/l_b^6.
\]
In the quasi-classical limit, $N\gg1$ and $n \approx N b /2\pi $ and above expression reduces to the result in the main text.

In general, when the Fermi surface deviates substantially from a perfect elliptical shape, one expects a multiplicity of cyclotron resonances. There are however certain symmetry constraints. For example, for a system with C$_n$ symmetry around the axis of the applied magnetic field, the spinon conductivity will only display resonances that evolve adiabatically from $(nk\pm 1)\omega_b$, where $k \in Z$ as the parameter that controls the anisotropy increases. Thus for a C$_6$ symmetric Fermi surface, such as that of the ideal Hubbard model in the triangular lattice, resonaces of the spinon conductivity are expected to be closer to $\omega_b, 5\omega_b, 7\omega_b,.....$. The corresponding physical conductivity of the spinon fermi surface state is expected to have resonances close to $5\omega_b, 7\omega_b$.

\begin{table*}[t]
	\begin{tabular}{ |c|c| c| c|c|c|c|c|}
		\hline
		Materials & ~$\omega_{\text{Mott}}$ (meV) &~$a$~(\AA)& ~$\gamma$ (mJ~mol$^{-1}$K$^{-2}$) &~ $m_\Psi$ ($m_e$) &~ $k_F$ (nm$^{-1}$)& ~$\varepsilon_F$ (meV) &~$ \omega_{b=1\text{T}}$(meV) \\ \hline
		YbB$_{12}$ (Ref.\onlinecite{Liu2018}) & 4~\cite{Batkova2006} & 7.5~\cite{Liu2018}&0.94 &3~\cite{Liu2018}& 0.96~\cite{Liu2018}& 12 & 0.04\\ \hline
		YbB$_{12}$ (Ref.\onlinecite{Xiang2018}) & 4~\cite{Batkova2006} & 7.5~\cite{Liu2018} & 3.8~\cite{Xiang2018} &6.7~\cite{Xiang2018} &1.6~\cite{Xiang2018} & 14 & 0.02\\ \hline
		SmB$_{6}$ ($\alpha$ sheet)& 2-5~\cite{Hartstein2017} & 4.1 ~\cite{Hartstein2017}& 2.21 &0.7 ~\cite{Hartstein2017} &4.85 ~\cite{Hartstein2017} & 1270&0.16\\ \hline
		SmB$_{6}$ ($\rho$ sheet)& 2-5~\cite{Hartstein2017}  & 4.1 ~\cite{Hartstein2017} & 0.47& 0.18~\cite{Hartstein2017} & 1.00~\cite{Hartstein2017} & 210&0.64\\ \hline
		SmB$_{6}$ ($\rho'$ sheet)& 2-5~\cite{Hartstein2017}   & 4.1 ~\cite{Hartstein2017}& 0.07&0.12~\cite{Hartstein2017} & 0.23~\cite{Hartstein2017} & 17&0.94\\ \hline
		1T-TaS$_2$ & 200~\cite{Qiao2017}   & 3.3~\cite{Tsen2015}& 0.1~\cite{He2018} &0.11& 2.25& 1810&1.1\\ \hline
		~EtMe$_3$Sb[Pd(dmit)$_2$]$_2$ & 80~\cite{Pustogow2018} & 6.4~\cite{Kato2012}& 19.9 ~\cite{Yamashita2011} & 11.3 & 4.2 & 59 & 0.01 \\ \hline
		~$\kappa$-(ET)$_2$Cu$_2$(CN)$_3$& 87 ~\cite{Pustogow2018}  &8.6~\cite{Komatsu1996}& 12~\cite{Yamashita2008} &3.8 & 3.13& 98 &0.03\\
		\hline
	\end{tabular}
	\caption{\label{Table:Estimates} Estimates of cyclotron energies for various candidate systems. Values measured in experiments are indicated by cited references.}
\end{table*}

\section{Estimates of transport life-time}\label{Appendix:Fluctuations}
Following the discussions in Ref.\onlinecite{Lee1992}, the imaginary parts of spinon self-energy in dimensions $d=2,3$ due to gauge field interactions at zero temperature are:
\begin{eqnarray}\label{Eqn:SelfEnergy}
\Sigma''_{d=2}=-\frac{ \nu_2 k_F}{2\pi m_\Psi^2} \int^{\omega}_0 d\omega \int_0^\infty \frac{\omega q/\gamma_2}{\omega^2+(\chi_d q^3/\gamma_2)^2} dq\\
\Sigma''_{d=3}= -\frac{ \nu_3 }{2 m_\Psi^2} \int^{\omega}_0 d\omega \int_0^\infty \frac{\omega q^2/\gamma_3}{\omega^2+(\chi_d q^3/\gamma_30)^2} dq,
\end{eqnarray}
where $\nu_d$ is density of state for one spin in $d$-dimensions. The constant $\gamma_d$ is defined by the $d$-dimensional transverse spinon conductivities $\sigma_d \sim i\gamma_d \omega/q$ in the $\omega, v_F q \ll \varepsilon_F$ limit. $\chi_d=\chi_d^\varphi+\chi_d^\Psi$ is the sum of chargon and spinon diamagnetic susceptibilities in $d$-dimensions. Note that $\chi_d$ is not the physical susceptibility which is given by a formula similar to Eqn.(\ref{Eqn:ILRule}) in the main text. To calculate transport life-time, the integrand in (\ref{Eqn:SelfEnergy}) is multiplied by $(q/k_F)^2/2$:
\begin{eqnarray}
\Gamma_{d=2}=-\frac{1}{8\pi^2 m_\Psi k_F\gamma_2} \int d\omega  \int \frac{\omega q^3}{\omega^2+(\chi q^3/\gamma_2)^2} dq,\\
\Gamma_{d=3}=-\frac{1}{8\pi^2 m_\Psi k_F\gamma_3} \int d\omega  \int \frac{\omega q^4}{\omega^2+(\chi q^3/\gamma_3)^2} dq.
\end{eqnarray}
Integration over $q$ and then $\omega$ gives the result:
\begin{equation}\label{Eqn:Width1}
\Gamma_{d=2}=\frac{\gamma_2^{1/3}}{32\sqrt{3}\pi m k_F\chi_2^{4/3}} \omega^{4/3},~\Gamma_{d=3}=\frac{\gamma_3^{2/3}}{40\pi m k_F\chi_3^{5/3}} \omega^{5/3}.
\end{equation}
Away from the Mott transition, the spinon contribution to $\gamma_{2,3}$ and $\chi_{2,3}$ is expected to be dominant~\cite{Lee1992}. Then it can be shown that $\gamma_2=k_F/\pi$, $\gamma_3=k_F^2/4\pi$, $\chi_2=1/12\pi m_\Psi$ and $\chi_3=k_F/12\pi^2m_\Psi$. Substituting them into Eqn.(\ref{Eqn:Width1}) we obtain the final result:
\begin{equation}\label{Eqn:Width2}
\Gamma_{d=2}= \frac{6^{1/3}\sqrt{3}}{8} \varepsilon_F^{-1/3} \omega^{4/3};~\Gamma_{d=3}= \frac{1}{5}\bigg(\frac{3\pi}{2}\bigg)^{5/3} \varepsilon_F^{-2/3} \omega^{5/3}.
\end{equation}
This has the same form as (\ref{Eqn:Width}) in the main text with $\Lambda = \varepsilon_F, C_2\approx0.4$ and $C_3\approx2.65$.

\section{Estimates for specific material candidates}\label{Appendix:Estimates}
In this section we present order-of-magnitude estimates of the energy scales relevant to spinon cyclotron resonance in several material candidates. Near the Mott transition, ratio between physical and gauge magnetic fields $\alpha \approx 1$~\cite{Inti2018}, and thus, for purposes of roughly estimating the cyclotron energy, we will assume the spinons to experience the full external magnetic field. In estimating, we have made use of the expression relating linear specific heat coefficient per mole of formula units $\gamma$ to density of states $\nu$:
\[
\gamma = \frac{\pi^2}{3} k_B^2 \nu \times \frac{N_A V_{\text{uc}}}{Z}.
\]
$N_A$ is the Avogadro constant, $V_{\text{uc}}$ is the unit cell volume and $Z$ number of formula units per unit cell, and for simplicity we take the spinons to have a parabolic dispersion. The measured value of $\gamma$ together with the assumption that the spinons form a fermi surface with one spinon per unit cell of a triangular lattice allow to estimate the effective parameters for 1T-TaS$_2$ and for the organic materials EtMe$_3$Sb[Pd(dmit)$_2$]$_2$ and	$\kappa$-(ET)$_2$Cu$_2$(CN)$_3$, shown in Table \ref{Table:Estimates}. The Mott scales are taken either from optical or tunneling spectroscopy according to the cited references. We remind the reader that the Mott gap in the organics can be conveniently tuned to zero through a metal to insulator transition with pressure in the organics.

For the three-dimensional mixed valence insulators we have taken $m_\Psi$ and $k_F$ directly from the quantum oscillation measurements. For YbB$_{12}$, there is a discrepancy between the values measured by two groups~\cite{Liu2018,Xiang2018}, and we list both of them. In calculating $\gamma$ for SmB$_6$, we use the fact that there are three inequivalent sheets of fermi ellipsoids with different band parameters~\cite{Hartstein2017}: for $\alpha$ sheet there are $3$ equivalent pockets, while for $\rho$ and $\rho'$ there are $12$ such pockets.

\bibliography{bib}

\end{document}